\documentstyle[11pt]{article}
\begin{document}

\centerline{\Large\bf A path-integral approach to the collisional}
\vskip 5pt
\centerline{\Large\bf Boltzmann gas}
\vskip 30pt

\centerline{C. Y. Chen}
\centerline{Dept. of Physics, Beijing University of Aeronautics}
\centerline{and Astronautics, Beijing 100083, PRC}
\vskip 20pt
\centerline{Email: cychen@public2.east.net.cn}

\vfill

\noindent{\bf Abstract}: Collisional effects are included in the 
path-integral 
formulation that was proposed in one of our previous paper for the
collisionless Boltzmann gas. In calculating the 
number of molecules entering a six-dimensional phase 
volume element due to collisions, 
both the colliding molecules and the scattered molecules are allowed to have
distributions; thus the calculation is done smoothly and no singularities
arise.

\vskip10pt
\noindent PACS number: 51.10.+y.

\newpage
\section{Introduction}
In our previous works, 
we proposed a path-integral approach to the collisionless Boltzmann 
gas\cite{chen}\cite{chen1}. 
It is assumed in the approach that there are continuous and discontinuous 
distribution functions in realistic Boltzmann gases: continuous distribution 
functions are produced by continuous distribution functions that exist 
previously and 
discontinuous distribution functions are caused by boundary effects. 
(Boundaries can block and reflect
molecules in such a way that distribution functions become  
discontinuous in the spatial space as well as in the velocity space.)
To treat these two kinds of distribution functions at the same time, 
a different type of distribution function, 
called the solid-angle-average distribution function, is introduced as 
\begin{equation}\label{fdef} f(t,{\bf r},v,\Delta\Omega)=\frac 
1{\Delta\Omega} \int f(t,{\bf r},{\bf v}) d\Omega, \end{equation}
where $\Delta\Omega$ represents one of the solid angle ranges in the 
velocity space defined by the investigator and $f(t,{\bf r},{\bf v})$ is the 
 ``ordinary'' distribution function. By letting each of
$\Delta\Omega$ be 
adequately small, the newly employed distribution function is capable of 
describing gas dynamics with any desired accuracy. 
Provided that collisions in a Boltzmann gas can be neglected, the 
solid-angle-average distribution function is found to be  
\begin{equation}\label{sur0}\begin{array}{r}\displaystyle 
f(t,{\bf r},v,\Delta\Omega)=\frac 1{\Delta \Omega}  \int_{\Delta S_1} \frac 
{f^{ct}(t_0,{\bf r}_0,v,\Omega_0)|\cos\alpha| dS_0}{ 
|{\bf r}-{\bf r}_0 |^2}U_{{\bf r}_0{\bf r}}\quad  \vspace{4pt} \\
\qquad\qquad\qquad\qquad
\displaystyle +\frac 1{\Delta\Omega}\int_{\Delta S_2} \frac {\eta(t_0,
{\bf r}_0,v,\Omega_0) dS_0}{|{\bf r}-{\bf r}_0|^2 v^3}U_{{\bf r}_0{\bf r}},
\end{array} \end{equation}
where, referring to Fig. 1, $\Delta S_1$ is an arbitrarily chosen virtual 
surface within the effective cone defined by $-\Delta\Omega$ at the point
${\bf r}$, $\Delta S_2$ stands for all boundary surfaces within the 
effective cone, $\eta$ is the local emission rate of boundary surface 
(acting like a 
surface-like molecular source), ${\bf r}_0$ represents the position of 
$dS_0$, $\Omega_0$ is the solid angle of
the velocity but takes the direction of $({\bf r}-{\bf r}_0)$, $t_0$ is the 
local time defined by $t_0=t-|{\bf r}-{\bf r}_0|/v$, $\alpha$ is the angle 
between the normal of $dS_0 $ and the vector ${\bf r}-
{\bf r}_0$, $f^{ct}$ is the continuous part of the distribution function
existing previously, and $U_{{\bf r}_0{\bf r}}$ is the path-clearness step 
function, which is equal to
1 if the path $\overline{{\bf r}_0{\bf r}}$ is free from blocking otherwise 
it is equal to 0.  

The objective of this paper is to include collisional effects in the 
path-integral formalism.

According to the conventional wisdom collisions can be analyzed by the 
method 
developed by Boltzmann long ago, in which it is understood that there is a 
symmetry between the ways molecules enter and leave a phase volume 
element. Peculiarly enough, this well-accepted understanding includes 
actually hidden fallacies\cite{chen2}, which can briefly be summarized as 
follows. In terms of studying collisions in a Boltzmann gas, there are two 
issues that are supposedly important. The first one is related to how many 
collisions take place within a phase volume element and during a certain 
time; the second one is related to how the scattered molecules will, 
after collisions, spread out over the velocity space and over the spatial 
space. These two issues involve different physics and have to be formulated 
differently. If the molecules leaving a phase volume element is of interest, 
one needs to take care of only the first issue; whereas if the molecules 
entering a volume element is of interest, one needs to concern oneself with 
both the issues aforementioned. This imparity simply suggests that the 
time-reversal symmetry, though indeed exists for a single collision between 
two molecules, cannot play a decisive role in studying 
collective effects of collision. 

In the present paper we formulate the collisional effects partly in an 
unconventional way.  In deriving how many molecules make collisions, the 
standard method is employed without much discussion; but, in
formulating how scattered molecules enter a six-dimensional phase volume 
element, which is an absolute
must for the purpose of this paper, a rather different and slightly 
sophisticated approach is introduced.

In sec. 2, general considerations concerning basic collisional process are 
given. It is pointed out that only the scattering cross section in the
center-of-mass frame is well defined and can be employed in our studies.
Sec. 3 gives a formula that describes how a molecule, when moving along its
path, will survive from collisions. Sec. 4 investigates how molecular
collisions create molecules that enter a specific phase volume element. In
the investigation, both the colliding molecules and the scattered molecules
 are allowed to have distributions. (Otherwise, singularities will arise,
as Ref. 3 reveals.) Sec. 5 includes all the collisional effects in a 
complete path-integral formulation. In Sec. 6, approximation methods 
are introduced to make the new formulation more calculable and an 
application of the method is demonstrated. Sec. 7 offers a brief summary.

Throughout this paper, to make our discussion as 
simple as possible, it is assumed that molecules of interest are all
identical, but distinguishable, perfectly rigid spheres and they move freely 
 when not making collisions.

\section{General considerations of collision}
Firstly, we recall general features of binary collisions in terms of 
classical mechanics. Consider two molecules: one is called molecule 1 and 
the other molecule 2.
Let ${\bf v}_1$ and ${\bf v}_2$ label their respective velocities before the 
collision. The 
center-of-mass velocity and the velocity of molecule 1 relative to the 
center-of-mass are before the collision
\begin{equation} {\bf c}=\frac 12 ({\bf v}_1+{\bf v}_2)\quad{\rm and}\quad
{\bf u}=\frac 12({\bf v}_1-{\bf v}_2).\end{equation}
Similarly, the center-of-mass velocity and the velocity of molecule 1 
relative to the center-of-mass are after the collision 
\begin{equation} \label{cvp}    {\bf c}^\prime=\frac 12
({\bf v}_1^\prime+{\bf v}_2^\prime)\quad{\rm and}\quad
{\bf u}^\prime=\frac 12({\bf v}_1^\prime-{\bf v}^\prime_2).\end{equation}
The conservation laws of classical mechanics tell us that
\begin{equation} {\bf c}={\bf c}^\prime \quad{\rm and}\quad 
|{\bf u}|=|{\bf u}^\prime|=u.\end{equation}
Fig. 2 schematically illustrates the geometrical relationship of these 
velocities. Note that, for the collision defined  as above
the final velocities of the two molecules,
such as ${\bf u}^\prime$, ${\bf v}_1^\prime$ and ${\bf v}_2^\prime$, cannot
be completely determined unless the impact parameter of the collision is 
specified at the very beginning\cite{landau}.

At this point, mention must be made of one misconcept in that the usual
derivation of the Boltzmann equation gets involved\cite{chen2}. In an 
attempt to invoke the time-reversal symmetry of molecular collision, the 
standard treatment in textbooks\cite{reif} defines the scattering cross  
section in the laboratory frame in such a way that  
\begin{equation}\label{def}
\bar\sigma({\bf v}_1,{\bf v}_2\rightarrow {\bf v}_1^\prime,
{\bf v}_2^\prime)d{\bf v}_1^\prime d{\bf v}_2^\prime  \end{equation}
represents the number of molecules per unit time (per unit flux of 
type 1 molecules incident upon a type 2 molecule) emerging after scattering
with respective final velocities between ${\bf v}_1^\prime$ and
${\bf v}_1^\prime +d{\bf v}_1^\prime$ and between ${\bf v}_2^\prime$ and
${\bf v}_2^\prime+ d{\bf v}_2^\prime$. 

If a close look at expression (\ref{def}) is taken, we may find that
the value of $\bar\sigma$ in it is ill-defined. As Fig. 2b clearly
shows, the molecules of type 1, namely the ones with the velocities
${\bf v}_1^\prime$ after the collisions, will spread out over a
two-dimensional surface in the velocity space (forming a spherical shell 
with diameter $2u$) rather than over a three-dimensional velocity volume as
suggested by the definition. Because of this seemingly small fault, the 
value of $\bar \sigma$ actually depends on the size and shape of
$d{\bf v}_1^\prime$ and cannot be treated as a uniquely defined quantity
theoretically and experimentally.

Another type of scattering cross section, which is elaborated nicely in 
textbooks of classical mechanics and suffers from no difficulty, is in terms
of the relative velocities $\bf u$ and ${\bf u}^\prime$, as shown in Fig. 3.
It is defined in such a way that the area element
\begin{equation}\label{def1}
dS= \sigma (\Omega_{{\bf u}^\prime}) d\Omega_{{\bf u}^\prime}\end{equation}
represents the number of molecules per unit time (per unit flux of 
type 1 molecules with the relative velocity ${\bf u}$ incident upon a type 2
molecule) emerging after scattering with the final relative velocity
${\bf u}^\prime$ pointing in a direction within
the solid angle range $d\Omega_{{\bf u}^\prime}$. 
Note that the definition (\ref{def1}), in which the center-of-mass velocity 
$\bf c$ or ${\bf c}^\prime$ becomes irrelevant, makes good sense in the 
center-of-mass frame rather than in the
laboratory frame. 

Before finishing this section we turn to discussing how collisions can 
generally affect the solid-angle-average distribution function $f(t,{\bf 
r},v,\Delta\Omega)$ defined by (\ref{fdef}).  
In view of that gas dynamics of the Boltzmann gas develops along
molecular paths, as shown by (\ref{sur0}), we believe that 
collisional effects should also be investigated 
and formulated in terms of molecular paths. 
Fig. 4 illustrates that there are two types of processes. On one hand, a
molecule that would reach ${\bf r}$ with the velocity ${\bf v}$ at time $t$
may suffer from a collision and become irrelevant to the distribution
function; on the other hand, an ``irrelevant'' molecule may collide
with another molecule and then become relevant.
It should be stressed again that there is no symmetry between the two types
of processes. For the first type of process, we only need to
investigate what happens to a single molecule. If a collision takes place 
with it, we know that the molecule will depart from its original path, which
is sufficient as far as our formulation is concerned.
For the second type of process, we need to know: (i) how many collisions 
take 
place within the effective cone; (ii) how the scattered molecules spread out 
over the phase space. 
As stressed in the introduction, the second issue is particularly essential 
because of that the distribution function is nothing but the molecular 
density per unit phase volume, in other words we must concern ourselves with 
the scattered
molecules ``around'' the phase point $({\bf r},{\bf v})$, rather than the 
scattered molecules ``at''
the phase point $({\bf r},{\bf v})$.

In the next two sections, we will formulate the two processes respectively. 
Since the collisions are assumed to take place in terms of classical
mechanics all the calculations can be done without analytical difficulty.

\section{The surviving probability}

Consider a molecule moving along
a spatial path where many other molecules make their own motions. 
If $P(\tau)$ denotes
the probability that the molecule survives a time $\tau$ without suffering
a collision and $w(\tau) d\tau$ denotes the probability that the molecule
makes a collision between time $\tau$ and time $\tau+d\tau$, we must have a
simple relation
\begin{equation} P(\tau+d\tau)-P(\tau)=-P(\tau)w(\tau)d\tau,\end{equation}
which yields
\begin{equation} \frac 1P\frac{dP}{d\tau} =-w(\tau) .\end{equation}
Therefore, the surviving probability associated with a molecule moving from
${\bf r}_0$ to ${\bf r}$ with the velocity ${\bf v}$ can be expressed 
formally by
\begin{equation}\label{sur} P({\bf r}_0,{\bf r};{\bf v})=\exp(-\int_l 
w(\tau) d\tau),\end{equation}
where $l$ represents the path along that the molecule will move if no
collision takes place. For the Boltzmann gas under consideration, whose
molecules are assumed to be free from forces except in collisions, the path
of a molecule is nothing but the
segment of straight line linking up the two points. 

The surviving probability defined by (\ref{sur}) can be evaluated by the
standard approach\cite{reif}. To make this paper complete, we include the
evaluation in what follows. Suppose that the molecule encounters
a molecular beam with the velocity ${\bf v}_1$ at the path element $dl$. 
In terms of the molecular beam, the molecule has the speed $2u=2|{\bf u}|$,
in which ${\bf u}=({\bf v}-{\bf v}_1)/2$, and it occupies the volume
with respect to the beam
\begin{equation}
2u  \sigma(\Omega_{{\bf u}^\prime}) d\Omega_{{\bf u}^\prime} d\tau,
\end{equation}
where $\sigma(\Omega_{{\bf u}^\prime})$ and $\Omega_{{\bf u}^\prime}$ are  
defined in (\ref{def1}) and illustrated by Fig. 3.
The molecular density of the colliding beam is 
\begin{equation}  f(\tau,{\bf r}_l,{\bf v}_1)d{\bf v}_1,\end{equation}
where ${\bf r}_l$ is the position of the path element $dl$.
Thus, the total collision probability can be written as
\begin{equation}\label{ww1} w d\tau =d\tau 
\int_{{\bf v}_1}\int_{\Omega_{{\bf u}^\prime}} 
2u f(\tau,{\bf r}_l,{\bf v}_1) \sigma(\Omega_{{\bf u}^\prime}) 
d \Omega_{{\bf u}^\prime} d{\bf v}_1.  \end{equation}
In terms of (\ref{ww1}), the surviving probability (\ref{sur}) becomes
\begin{equation}\label{prr}
P({\bf r}_0,{\bf r};{\bf v})=\exp\left[ -\int_l \int_{{\bf v}_1}
\int_{\Omega_{{\bf u}^\prime}} 2u \sigma(\Omega_{{\bf u}^\prime}) 
f(\tau,{\bf r}_l,{\bf v}_1) 
d\Omega_{{\bf u}^\prime} d{\bf v}_1 d\tau. \right],\end{equation}
where $d\tau$ is the time period during that the molecule passes the
path element $dl$. 

The formula (\ref{prr}) describes how collisions make the number of 
molecules along a certain path decrease. The method employed has nothing
particularly new in comparison with that employed by the textbook treatment.
Before changing our subject, one thing worth mentioning. In deriving
(\ref{prr}), we had luck not to be concerned with how the scattered
molecules spread out over the phase space. It is readily understandable
that the same luck will not be there in the next section.

\section{The creation probability}
We now study the process in which collisions make molecules
give contributions to the solid-angle-average distribution 
function $f(t,{\bf r},v,\Delta \Omega)$. 

It should be mentioned that
in this section, unlike in the last sections, 
${\bf v}^\prime$ and ${\bf v}_1^\prime$ represent the velocities of 
colliding molecules while ${\bf v}$ and ${\bf v}_1$ represent
the velocities of scattered molecules. 

The see how the scattered 
molecules spread out over the  six-dimensional phase space, we consider 
a relatively small six-dimensional volume element as
\begin{equation}\label{pv} 
\Delta {\bf r}\cdot \Delta {\bf v} = \Delta {\bf r}\cdot v^2 \Delta v \Delta
\Omega.\end{equation}
In (\ref{pv}) $\Delta {\bf r}$ is chosen to enclose the point ${\bf r}$ in
$f(t,{\bf r},v,\Delta \Omega)$, $\Delta v$ to enclose the speed $v$ in
$f(t,{\bf r},v,\Delta \Omega)$;
and $\Delta \Omega$ is just the finite velocity solid-angle-range  
$\Delta \Omega$ in $f(t,{\bf r},v,\Delta \Omega)$. The discussion below will 
 be focused on molecules that really enter,  after collisions, this 
six-dimensional volume element.   

Note that in Fig. 5a each point in the spatial volume $\Delta{\bf r}$ 
defines an effective cone, within which physical events may make
impact on the distribution function at the point.
This means that the entire effective cone, with respect to the spatial 
volume 
element $\Delta {\bf r}$, must be slightly larger than the one defined 
solely  by ${\bf r}$, as shown by Fig. 5b.
Fortunately, we will, in the end of formulation, let 
\begin{equation}\label{rb} \Delta {\bf r}\rightarrow 0 \end{equation}
and thus the actual entire effective cone is only academically larger. On 
this understanding, we will not distinguish between the entire effective 
cone and the effective cone defined solely by the single point ${\bf r}$.

Look at molecular collisions taking place within the effective
cone shown in Fig. 6. 
Note that Fig. 6a, while coming to one's mind immediately, 
is not an appropriate picture to manifest the collision process 
affecting the distribution function at the point ${\bf r}$ since
 both the colliding molecules and scattered molecules in it
have no true distributions. 
(Ref. 3 analyzes the situation and brings out that 
such mental picture will finally lead to singularities.) 
In Fig. 6b, the velocity distributions of all colliding molecules and
scattered molecules are explicitly illustrated. Our task here is to 
formulate the relationship between the distribution functions of colliding 
molecules and the scattered molecules
(including their velocity distributions and spatial distributions).

We divide the entire effective cone into many individual regions, denoted by 
 $(\Delta {\bf r}_0)_i$. It is obvious that within each of the regions
collisions can generate
a certain number of molecules that will finally enter the phase volume 
defined by (\ref{pv}). 
Let $n^{cl}_i$ denote the number of such molecules. 
In what immediately follows, 
it is assumed that the generated molecules suffer no further collisions.   
The entire contributions of all collisions to the 
distribution function $f(t,{\bf r},v,\Delta \Omega)$ can then be expressed 
by \begin{equation}\label{fcl} f^{cl}(t,{\bf r},v,\Delta\Omega)=
\frac 1{\Delta {\bf r} \cdot \Delta {\bf v}}\sum\limits_i n^{cl}_i,
\end{equation}
in which $i$ runs all the divided regions within the effective cone. 
For later use, we wish to rewrite (\ref{fcl}) as
\begin{equation}\label{fcl1} f^{cl}(t,{\bf r},v,\Delta\Omega)
=\frac 1{\Delta\Omega}\sum\limits_i \frac{n^{cl}_i}{v^2(\Delta {\bf r})_i
 (\Delta v)_i}. \end{equation}
The advantage of (\ref{fcl1}) over (\ref{fcl}) is that 
$(\Delta {\bf r})_i$ and $(\Delta v)_i$ in (\ref{fcl1}) may be chosen to be 
different for different  $i$ as 
long as the molecular number of $n^{cl}_i$ is counted up accordingly. 
(Of course, all velocity 
directions of the molecules have to be within the solid angle range $\Delta
 \Omega$.)  

Now, consider a small, much smaller than $\Delta\Omega$, solid angle range 
$\Delta\Omega_0$ at a point ${\bf r}_0$ towards the point ${\bf r}$, as 
shown 
in Fig. 7a (${\bf r}_0$ is within the effective cone). 
It is easy to see that if collisions take place at ${\bf r}_0$,
the scattered molecules having velocities within $\Delta \Omega_0$ will 
spread out over the spatial volume element  
\begin{equation}\label{rv} \Delta{\bf r}\approx |{\bf r}-{\bf r}_0|^2 v 
\Delta \Omega_0 \Delta t,\end{equation}
as shown in Fig. 7b. Accordingly, they will spread out over the velocity 
volume element
\begin{equation}\label{vv} \Delta{\bf v}\approx  v^2 \Delta v 
\Delta\Omega_0, \end{equation}
as shown in Fig. 7c. 
Since $\Delta\Omega_0$ is much smaller than $\Delta\Omega$ (the
latter one is finite), a molecule having a velocity within $\Delta\Omega_0$
can be regarded as one having a velocity within $\Delta\Omega$. Thus, by
letting $(\Delta{\bf r})_i$ in (\ref{fcl1}) be equal to $\Delta{\bf r}$ of
(\ref{rv}) and letting $(\Delta v)_i$ in (\ref{fcl1}) be equal to
$\Delta v$ in (\ref{vv}), expression (\ref{fcl1}) becomes  
\begin{equation}\label{fcl2} f^{cl}(t,{\bf r},v,\Delta\Omega)
=\frac 1{\Delta\Omega}\int  \lim\limits_{(\Delta\Omega_0,
\Delta v,\Delta t)\rightarrow {(0,0,0)}} \frac{\rho^{cl}({\bf r}_0) 
d{\bf r}_0 }{v^3 \Delta v |{\bf r}-{\bf r}_0|^2 \Delta\Omega_0 \Delta t  }, 
\end{equation} 
where the integral is over the entire effective cone defined by 
${\bf r}$ and $-\Delta\Omega$ and $\rho^{cl}$ is the local density 
(per unit spatial volume) of the molecules  that are ``emitted'' from 
 ${\bf r}_0$ due to collisions and finally enter the speed range $\Delta v$
and the solid-angle range $\Delta\Omega_0$ during the time $\Delta t$. 

To determine the density $\rho^{cl}$, we have two tasks. One is to derive 
the collision rate at ${\bf r}_0$ and the other is to derive what fraction 
of the scattered molecules emerge with velocities within the range $\Delta v 
 \Delta\Omega_0$. 
The first task can be accomplished in a well-known way while the second one 
cannot. 

As discussed in the last section, 
a specific molecule with the initial velocity ${\bf v}^\prime$ that  
collides with a beam of molecules with the initial velocity
${\bf v}_1^\prime$ occupies a volume with respect to the beam
\begin{equation} 2 u \Delta t \sigma(\Omega_{\bf u}) d \Omega_{\bf u},
\end{equation}
where $\Omega_{\bf u}$ is the solid angle of the scattered relative velocity 
 ${\bf u}=({\bf v}-{\bf v}_1)/2$. The number of ``such specific'' molecules
within $d{\bf v}^\prime d{\bf r}_0$ is
\begin{equation}
f(t_0,{\bf r}_0,{\bf v}^\prime) d{\bf v}^\prime d{\bf r}_0.\end{equation}
The density of the molecules with ${\bf v}^\prime_1$ is characterized by
\begin{equation}
f(t_0,{\bf r}_0,{\bf v}_1^\prime) d{\bf v}_1^\prime. \end{equation}
Therefore, the number of all collisions within the spatial volume $d{\bf 
r}_0$ within the time $\Delta t$ is
\begin{equation}\label{num}
 d{\bf r}_0  \int d{\bf v}^\prime \int
 d{\bf v}_1^\prime \int d\Omega_{\bf u} f({\bf v}^\prime)
f({\bf v}_1^\prime) 2u \sigma(\Omega_{\bf u})\Delta t. \end{equation}

We now to evaluate the probability that the molecules expressed by 
(\ref{num}) 
enter the velocity range $\Delta v \Delta\Omega_0$.
Note that the integration of $d{\bf v}^\prime d{\bf v}_1^\prime$ is carried
out in the laboratory frame 
while the integration of $d\Omega_{\bf u}$ is in the center-of-mass frame, 
which makes the evaluation quite difficult. For this reason,
we make the integration conversion as
\begin{equation}\label{intconver} \int d{\bf v}^\prime \int d{\bf 
v}^\prime_1 \cdots =
\int d{\bf c}^\prime \int d\Omega_{{\bf u}^\prime} \int u^2 du \|J\| \cdots 
,\end{equation}
where $u=u^\prime$ is understood and $\|J\|$ represents the Jacobian
 between the center-of-mass frame and the laboratory frame
\begin{equation} \|J\|=\frac
{\partial({\bf v}^\prime,{\bf v}^\prime_1)}
{\partial({\bf c}^\prime,{\bf u}^\prime)}.\end{equation}
Equation (\ref{cvp}) tells us that the Jacobian is equal to $8$.

By making use of (\ref{num}) and (\ref{intconver}), the distribution 
function   
$f^{cl}$ expressed by (\ref{fcl2}) becomes
\begin{equation}\label{tem1} \begin{array}{l} \displaystyle
f^{cl}(t,{\bf r},v,\Delta \Omega)\approx \frac 1{\Delta\Omega} \int_{-\Delta
 \Omega} d{\bf r}_0 \int d{\bf c}^\prime \int d\Omega_{{\bf u}^\prime}
\int\int_{\Delta v \Delta\Omega_0} u^2d\Omega_{\bf u} du \vspace{4pt} \\
\displaystyle\qquad\quad\cdot
 \frac{\|J\|}{v^3 \Delta v |{\bf r}-{\bf r}^\prime|^2 \Delta\Omega_0}2u
\sigma(\Omega_{\bf u})f(t_0,{\bf r}_0,{\bf c}^\prime-{\bf u}^\prime)
 f(t_0,{\bf r}_0,{\bf c}^\prime+{\bf u}^\prime). \end{array} \end{equation}

In regard to expression (\ref{tem1}), some observations are made.
As mentioned in Sec. 2, if 
two molecular beams with definite velocities ${\bf v}^\prime$ and
${\bf v}_1^\prime$ collide with each other the scattered molecules will 
spread out only over a two-dimensional spherical surface in the velocity 
space, which implies that difficulty arises if the velocity distributions
of scattered molecules are of concern.
Whereas, in this expression, all the colliding molecules are allowed to have 
 distributions, the value of $u^\prime=u$ is allowed to vary and therefore 
the scattered molecules explicitly spread out over the velocity space
(as well as over the spatial space). 
Furthermore, by using the notation
\begin{equation} \int\int_{\Delta v\Delta\Omega_0} \cdots\cdots,
\end{equation}
we have ensured that only the scattered molecules of 
relevance are taken into account.

In Fig. 8, which is drawn for scattered molecules in the velocity space, 
we are concerned only with molecules that finally
enter the range $\Delta v\Delta\Omega_0 $. Allowing ${\bf u}$   
to vary a little bit, we may let the scattered molecules fill out 
the velocity range. Namely, we have 
\begin{equation}\label{vcell}\int\int_{\Delta\Omega_0\Delta v}  
u^2 d\Omega_{\bf u} du (\cdots)
 \approx  v^2 \Delta v \Delta \Omega_0  (\cdots),\end{equation}
where $(\cdots)$ represents other factors that have been treated as 
constants in terms of the infinitesimally small range of $\Delta v \Delta 
\Omega_0$.  

Inserting (\ref{vcell}) into (\ref{tem1}) and taking the limits 
$\Delta\Omega_0\rightarrow 0$ and $\Delta v\rightarrow 0$, we finally 
arrive at
\begin{equation} \label{coll} \begin{array}{l} \displaystyle
 f^{cl}(t,{\bf r},v,\Delta\Omega)=\frac 1{v\Delta \Omega} 
\int_{-\Delta \Omega} d{\bf r}_0 \int d{\bf c}^\prime \int d\Omega_{{\bf
u}^\prime}
\\   \displaystyle \quad\qquad \frac { \|J\|}{ |{\bf r}-{\bf r}_0|^2 }   
2 u \sigma(\Omega_{\bf u}) f(t_0,{\bf r}_0,{\bf c}^\prime-{\bf u}^\prime)
 f(t_0,{\bf r}_0,{\bf c}^\prime+{\bf u}^\prime) 
, \end{array} \end{equation}
where  $t_0=t-{|{\bf r}-{\bf r}_0|}/ v$ and the integration 
\begin{equation} \int d\Omega_{{\bf u}^\prime}\cdots  \end{equation}
is over the entire solid angle ($0\rightarrow 4\pi$). Note that $u$, ${\bf
u}^\prime$ and ${\bf u}$ in the integrand of (\ref{coll}) have to be
determined skillfully. First use ${\bf c}={\bf c}^\prime$ and
${\bf v}=v({\bf r}-{\bf r}_0)/|{\bf r}-{\bf r}_0|$ to determine ${\bf u}$,
then use $u=|{\bf u}|$ and
$\Omega_{{\bf u}^\prime}$ to determine ${\bf u}^\prime$, as shown in Fig. 9.

We have directly 
formulated the contribution to the solid-angle-average distribution function
from collisions. It should be noted that the formulation
can be done only under the condition that the velocity solid-angle range
$\Delta \Omega$ is finite:
if both $\Delta\Omega$ and $\Delta \Omega_0$ in the formulation were assumed 
to be
infinitesimally small, the limiting processes concerning the two quantities
would not be in harmony with each other. This shows again that the
introduction of the solid-angle-average distribution function is a must to
the gas dynamics of Boltzmann gas.
 
\section{Complete formulation }
The complete formulation for the collisional Boltzmann gas is now in order. 
In Fig. 10, we have depicted 
a piece of boundary and some collisions taking place within  the
effective cone. As said before, all these events  
can directly contribute to the solid-angle-average distribution function.

We then use the following sum to represent the total distribution function
\begin{equation}\label{tf} 
f(t,{\bf r},v,\Delta\Omega)= f_{(i)}+f_{(ii)}+f_{(iii)}, \end{equation}
where $f_{(i)}$, $f_{(ii})$ and $f_{(iii)}$ represent the 
contributions from the existing continuous distribution function, from
the piece of boundary and from the collisions respectively. 
For simplicity, no other types of distribution functions are assumed to 
exist within the effective cone.

As has been illustrated in Sec. 3, a molecule that makes its motion towards
the point ${\bf r}$ may suffer a collision with other molecules. The 
involved surviving probability $P$ has been defined by (\ref{prr}). 
By taking the probability into account, the first term in (\ref{sur0}) 
becomes 
\begin{equation} f_{(i)}(t,{\bf r},v,\Delta\Omega)
=\frac 1{\Delta \Omega} \int_{\Delta S_1} \frac{          
f^{ct} (t_0,{\bf r}_0,v,\Omega_0) |\cos\alpha| dS_0}
{|{\bf r}-{\bf r}_0|^2}   P({\bf r}_0,{\bf r},{\bf v}_0),\end{equation}
where ${\bf v}_0$ in $P$ take the value of $v$ and points to the direction 
of 
$({\bf r}-{\bf r}_0)$. 

By taking the same effect into account, the second term in (\ref{sur0})
can be expressed by   
\begin{equation}\label{ffi} f_{(ii)}(t,{\bf r},v,\Delta\Omega) =
\frac 1{\Delta \Omega} \int_{\Delta S_2}
\frac {\eta(t_0,{\bf r}_0,v,\Omega_0) d S_0}  
{|{\bf r}-{\bf r}_0|^2v^3} P({\bf r}_0,{\bf r},{\bf v}_0) .\end{equation}

A rather detailed discussion on the molecular emission rate $\eta$ has been
included in Ref. 1. Here, we content ourselves with pointing out that the
rate $\eta$ satisfies 
the normalization condition at the surface element $d S_0$. If no molecular
absorption and production by the surface element are assumed, the following
expression holds
\begin{equation} \int \eta(t,{\bf r},v_1,\Omega_1) d\Omega_1 dv_1 = 
\int v_2f(t,{\bf r},v_2,\Omega_2)|\cos\theta| v_2^2 d\Omega_2 dv_2
,\end{equation}
in which $\theta$ is the angle between the velocity ${\bf v}_2$ and the 
normal  of $dS_0$, $\Omega_1$ points to an outward direction of 
$d S_0$ and $\Omega_2$ points to an inward direction of $d S_0$. The 
concrete relation between $\eta(t,{\bf r},{\bf v})$ and $f(t,{\bf r},{\bf 
v})$ must, of
course, be ultimately determined by experimental data\cite{kogan}. 
                                     
In obtaining (\ref{coll}) for the distribution function created by 
collisions, further collisions were excluded. To include 
possible further collisions, 
the contribution expressed by (\ref{coll}) needs to be modified as
\begin{equation}\label{collf}\begin{array}{l} 
f_{(iii)}(t,{\bf r},v,\Delta\Omega)= 
\displaystyle\frac 1{v\Delta \Omega} \int_{-\Delta \Omega} d{\bf r}_0 
\int d{\bf c}^\prime\int d\Omega_{{\bf u}^\prime}\\ \displaystyle
\quad\quad \frac {\|J\|} { |{\bf r}-{\bf r}_0|^2 }   2u \sigma
(\Omega_{{\bf u}^\prime}) f(t_0,{\bf r}_0,{\bf c}^\prime-{\bf u}^\prime) 
 f(t_0,{\bf r}_0,{\bf c}^\prime+{\bf u}^\prime) 
 P({\bf r}_0,{\bf r},{\bf v}_0) , \end{array} \end{equation}
where the integration of $d{\bf r}_0$ is over the entire effective cone
defined by ${\bf r}$ and $-\Delta\Omega$,
$|{\bf v}_0|\equiv v$ and takes the direction of $({\bf r}-{\bf r}_0)$, 
${\bf u}^\prime$ is defined by $u=|{\bf c}-{\bf v}_0|$ and 
$\Omega_{{\bf u}^\prime}$.

In these formulas, the probability $P$ should be set to be zero at the very 
beginning if there is physical blocking along the path 
$\overline{{\bf r}_0{\bf r}}$. 

Equations (\ref{tf})-(\ref{collf}) constitute a complete set of integral
equations that describe the collisional Boltzmann gas defined in this paper.
The formulation proves in a theoretical way an obvious intuition that the 
distribution function at a
spatial point can directly be affected by physical events taking place 
at other, even remote, points 
in view of the fact that 
 a molecule can freely pass any distance in a certain probability.
In this sense, the picture here is more ``kinetic'' than that associated 
with 
the Boltzmann equation, in which physical events have to make their 
influence region by region (like what happen in a continuous medium).  

Another comment is about the famous $H$-theorem. 
If the involved distribution function is initially nonuniform in the spatial
space and non-Maxwellian in the velocity space, the resultant distribution
function given by the formalism will approach the uniform Maxwellian. Though
such explicit proof has not been achieved yet, we believe that this must be
the case by noticing a general discussion stating that as long as a
statistical process is a Markoffian
one the $H$-theorem must hold true\cite{rave}.

\section{Approximation and application} 
Although the formulation offered in the previous section is formally
complete, there still exist difficulties that hinder one from performing
calculation for a realistic gas. Unlike the solution for the collisionless
Boltzmann
gas, given by (\ref{sur0}), the equation system in the last section, namely
(\ref{tf})-(\ref{collf}), is an integral-equation set. Without knowing
the entire history of the distribution function $f(t)$, the integrals in
the system cannot be evaluated accurately.

Fortunately, there are situations for which adequate approximations can
be introduced and meaningful results can be obtained. In what follows, we 
first deal with weakly collisional gases and then give some discussion on 
how the consideration can apply to more general cases.

If the density of a Boltzmann gas is relatively low, 
by which we imply that the mean free path of molecules is not too short 
comparing with the length scale of the system or that the mean free time is 
not too short
comparing with the time scale of the phenomena of interest, we may 
apply the following iterating procedure to calculate the distribution
function. 

Firstly, we assume that the system can be treated as a collisionless 
Boltzmann gas and the 
collisionless formulation can directly applied. 
Namely, we have the zeroth-order solution 
\begin{eqnarray} 
f^{[0]}_{(i)}=\frac 1{\Delta \Omega} \int_{\Delta S_1} \frac{          
f^{ct} (t_0,{\bf r}_0,v,\Omega_0) |\cos\alpha| dS_0}
{|{\bf r}-{\bf r}_0|^2}\\ 
  f^{[0]}_{(ii)} =
\frac 1{v^3 \Delta \Omega} \int_{\Delta S_2}
\frac {\eta(t_0,{\bf r}_0,v,\Omega_0) d S_0}  
{|{\bf r}-{\bf r}_0|^2}  . \end{eqnarray}

Then, we can construct the first-order distribution function
by inserting the zeroth-order solution into all right sides of the equations 
  (\ref{tf})-(\ref{collf}), which yields 
\begin{eqnarray} 
f^{[1]}_{(i)}=\frac 1{\Delta \Omega} \int_{\Delta S_1} \frac{          
f^{ct} (t_0,{\bf r}_0,v,\Omega_0) |\cos\alpha| dS_0}
{|{\bf r}-{\bf r}_0|^2}P^{[0]}({\bf r}_0,{\bf r},{\bf v}_0) \\
  f^{[1]}_{(ii)} =
\frac 1{v^3 \Delta \Omega} \int_{\Delta S_2}
\frac {\eta(t_0,{\bf r}_0,v,\Omega_0) d S_0}  
{|{\bf r}-{\bf r}_0|^2}P^{[0]}({\bf r}_0,{\bf r},{\bf v}_0) 
 . \end{eqnarray}
and
\begin{equation}\label{coll1}\begin{array}{l}\displaystyle
 f^{[1]}_{(iii)} = \frac 1{ v\Delta \Omega} \int_{-\Delta \Omega} d{\bf r}_0 
 \int d{\bf c}^\prime \int d\Omega_{{\bf u}^\prime} \frac{\|J\|}{|{\bf r}-
{\bf r}_0|^2} \vspace{4pt} \\
\displaystyle \qquad 2u\sigma(\Omega_{\bf u})   
 f^{[0]}(t_0,{\bf r}_0,{\bf c}^\prime-{\bf u}^\prime)
 f^{[0]}(t_0,{\bf r}_0,{\bf c}^\prime+{\bf u}^\prime) 
 P^{[0]}({\bf r}_0,{\bf r},{\bf v}_0) . \end{array} \end{equation}
In all the first-order formulas, the surviving probability is defined as
\begin{equation}\label{ptt} P^{[0]}({\bf r}_0,{\bf r};{\bf v}_0)=\exp\left[
-\int_l \int_{{\bf v}_1} \int_{\Omega_{{\bf u}^\prime}} 2u
\sigma(\Omega_{{\bf u}^\prime}) f^{[0]}(\tau,{\bf r}_l,{\bf v}_1) 
d\Omega_{{\bf u}^\prime} d{\bf v}_1 d\tau. \right].\end{equation}
In equations (\ref{coll1}) and
(\ref{ptt}), $f^{[0]}$ is the total zeroth-order
distribution function, namely $f^{[0]}=f_{(i)}^{[0]}+f_{(ii)}^{[0]}$.

Along this line, we can formulate higher-order solutions for dilute gases. 
If the gas of interest is rather dense, the approximation method presented 
above 
may not work effectively. One wishes, however, to point out that for the
regions near boundaries, where the distribution function suffers from most
irregularities and collisions between molecules have no enough time to erase
such irregularities, the introduced method should still make sense.
In view of this, it is expected that a hybrid method will be developed,
in which the approach here and other effective approaches, such as
the ordinary fluid theory, can be
combined into one scheme so that more practical gases become treatable.

To illustrate the application of our approximation scheme, we investigate a
gas leaking out of box through a small hole. For simplicity, we assume,
referring Fig. 11, that the zeroth-order solution of the leaking gas 
is confined to a ``one-dimensional thin pipe'' (shaded in the figure), which
 can be expressed by 
\begin{equation}\label{zero} f^{[0]}=\left\{ \begin{array}{ll}
C_0\exp\left[- {mv_x^2}/(2\kappa T)\right]&({\rm inside\; the \; pipe })\\
0 &  ({\rm outside\; the \; pipe })
\end{array}\right.\end{equation}
and then we try to determine the collisional effects of the distribution 
function.

Note that the distribution function expressed by (\ref{zero}) 
is kind of special  
so that the formula (\ref{coll1}) should be slightly modified. For this
purpose, we write the differential collision probability  as
(the subindex $x$ of $v_x$ is suppressed)
\begin{equation}
[f(v^\prime) \Delta S dx_0 dv^\prime][ f(v_1^\prime) dv_1^\prime ]
[2u \sigma(\Omega_{\bf u}) d\Omega_{\bf u}],\end{equation}
where $\Delta S$ is the cross area of the pipe. 
By making the variable transformation, we obtain 
\begin{eqnarray*}
\int_0^{\infty} dv^\prime \int_0^{\infty} dv_1^\prime \cdots
=\int_0^{\infty} dc^\prime \int_{-\infty}^{+\infty} du^\prime \|J\|\cdots\\ 
=\int_0^{\infty} dc^\prime \int_0^{\infty} u^2 du  4 (u^2)^{-1}\cdots.
\end{eqnarray*}
In a way similar to that has been presented in the last section,
 we finally arrive at
\begin{equation}\label{pra}
 f^{[1]}[(\Delta\theta)_i]=\frac{\Delta S}{ v (\Delta \theta)_i}
\int\limits_{-(\Delta\theta)_i} d x_0 \int\limits_0^\infty dc^\prime \frac
{8\sigma(\Omega_{\bf u})}{u|{\bf r}-{\bf r}_0|^2} f^{[0]}(c^\prime+u^\prime)
f^{[0]}(c^\prime-u^\prime),\end{equation}
where $(\Delta \theta)_i$ is the polar angle range set by the investigator
(the azimuthal angle range is irrelevant in the case).

The formula (\ref{pra}) can be calculated easily with a computer. 
Referring to Fig. 11b, we set 
$$ v=1,\quad r_\perp =1,\quad \frac m{2\kappa T}=1, $$
let $(\Delta \theta)_i$ be the interval
$$ \left[0.4\pi-0.52+\frac i{50},0.4\pi-0.5+\frac{i}{50}\right]
$$
and notice $\sigma(\Omega_{\bf u})$ is constant\cite{landau}.
The numerical results are listed as the following:
(normalized by $f[(\Delta\theta)_0]$)
\begin{equation} \begin{array}{l}
f[(\Delta\theta)_0]=1.00000e+00  \\
f[(\Delta\theta)_{5}]=5.23910e-01     \\
f[(\Delta\theta)_{10}]=2.01786e-01     \\
f[(\Delta\theta)_{15}]=5.03200e-02       \\
f[(\Delta\theta)_{20}]=4.76340e-03   \\
f[(\Delta\theta)_{25}]=5.11013e-05 
 .\end{array} \end{equation} 

\section{Summary}

In this paper, we have proposed a complete mathematical scheme to deal with
the Boltzmann gas. The scheme has many new features. In addition to
those given in Ref. 1, some related to treating collisional effects 
are the following.

Firstly, 
collisional effects are investigated in the full velocity-and-position 
space. 
In particular, a six-dimensional volume element is explicitly defined 
and a calculation concerning molecules entering 
the volume element is directly performed.

Secondly, both the colliding molecules and scattered molecules are allowed 
to have distributions. In other words, we consider the full and collective
behavior of collisions, in which the time-reversal symmetry existing for a
collision of two molecules plays almost no role.

Thirdly,
the treatment in this approach is consistent with the previous approach
to the collisionless Boltzmann gas in the sense that all the formulas are
given in terms of what happen along molecular paths.

Finally, the resultant formulas of this approach are, in many practical
situations, calculable by means of today's computer.

It is believed that this approach will be developed further so that  
a better understanding of complicated fluid phenomena can be achieved.

\newpage
\centerline{\LARGE \bf Figure captions}

\begin{enumerate}

\item A physical surface and a virtual surface within the effective cone
defined by ${\bf r}$ and $\Delta\Omega$.
\item
A collision between two molecules. (a) The molecular velocities before the 
collision. (b) The molecular velocities after the collision.

\item 
The scattering cross section in the center-of-mass frame.
(a) Solid angles and relative velocities.
(b) The relation between the cross section and solid angle range. 
 
\item Two types of collision processes.

\item Effective cones. (a) For a single spatial point. (b) For a given 
spatial volume.

\item  (a) A mental picture in 
which two molecular beams collide with each other. (b) A mental picture in 
which both colliding molecules and scattered molecules have distributions.

\item (a) A solid angle range $\Delta\Omega_0$ towards the point ${\bf r}$.
(b) The distribution of scattered molecules
in the spatial space. (c) The distribution of scattered molecules in the 
velocity space.

\item The velocity distribution of scattered molecules in the center-of-mass 
 frame and in the laboratory frame.

\item Relations between various essential vectors in the formulation. 

\item Contribution to the solid-angle-average distribution function from 
different sources.
\item
A gas leaking out of a container through a small hole. 

\end{enumerate}

\newpage

\noindent {\bf Figure 1 }

\setlength{\unitlength}{0.014in} 
\begin{picture}(200,140)

\put(59,42){\makebox(10,8)[l]{${\bf r}$}}
\put(154.97,52.49){\circle*{1}}
\put(154.87,54.97){\circle*{1}}
\put(154.71,57.45){\circle*{1}}
\put(154.48,59.93){\circle*{1}}
\put(154.19,62.40){\circle*{1}}
\put(153.83,64.86){\circle*{1}}
\put(153.41,67.31){\circle*{1}}
\put(152.92,69.75){\circle*{1}}
\put(152.38,72.18){\circle*{1}}
\put(151.76,74.59){\circle*{1}}
\put(151.09,76.98){\circle*{1}}
\put(150.35,79.36){\circle*{1}}
\put(149.55,81.71){\circle*{1}}
\put(148.69,84.04){\circle*{1}}
\put(147.77,86.35){\circle*{1}}
\put(147.77,86.35){\circle*{1}}
\put(146.64,88.72){\circle*{1}}
\put(145.27,90.95){\circle*{1}}
\put(143.68,93.02){\circle*{1}}
\put(141.88,94.92){\circle*{1}}
\put(139.89,96.62){\circle*{1}}
\put(137.73,98.10){\circle*{1}}
\put(135.43,99.35){\circle*{1}}
\put(133.02,100.35){\circle*{1}}

\put(60,52){\line(1,0){60}}
\put(140,52){\line(1,0){30}}
\put(130,50){\oval(20,40){}}
\put(60,52){\line(3,2){90}}
\put(60,52){\line(-1,0){20}}
\put(60,52){\line(-3,-2){16}}
\put(26,42){\makebox(10,8)[l]{$\Delta\Omega$}}
\put(45,68){\makebox(10,8)[l]{$-\Delta\Omega$}}
\put(103.5,59){\makebox(10,8)[c]{$\Delta S_2$}}
\put(148,95){\makebox(10,8)[c]{$\Delta S_1$}}
\put(72,56){\line(-1,1){10}}

\end{picture}

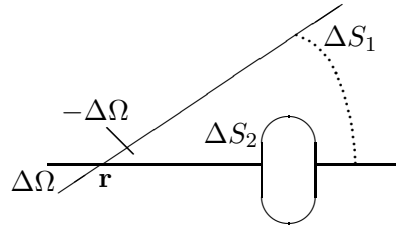
\noindent {\bf Figure 2}

\setlength{\unitlength}{0.013in}
\begin{picture}(200,160)


\put(63,15){\makebox(35,8)[l]{\bf (a)}}
\put(233,15){\makebox(35,8)[l]{\bf (b)}}

\put(168,85){\line(1,0){90}}
\put(233,85.5){\vector(1,0){1}}
\put(168,85){\line(3,1){108}}
\put(168,85){\vector(3,1){62}}
\put(258,85){\vector(1,2){12}}
\put(258,85){\line(1,2){18}}
\multiput(258,85)(-1.5,-3){12}{\circle*{1.2}}
\put(222,57.5){\vector(2,-1){1}}
\put(168,85){\line(2,-1){71.5}}

\put(108,85.5){\vector(1,0){1}}
\put(53,85){\line(1,0){90}}
\put(65.5,58){\vector(1,-2){1}}
\put(53,85){\line(1,-2){18}}
\put(35,121){\line(3,-1){108}}
\put(109,96){\vector(3,-1){1}}
\multiput(53,85)(-1.5,3){12}{\circle*{1.2}}
\put(71,48.5){\line(2,1){72}}
\put(71,48.5){\vector(2,1){42}}

\put(209,106){\makebox(35,8)[l]{${\bf v}_1^\prime$}}
\put(205,49){\makebox(35,8)[l]{${\bf v}_2^\prime$}}
\put(233,88.5){\makebox(35,8)[l]{${\bf c}^\prime$}}
\put(268,95){\makebox(35,8)[l]{${\bf u}^\prime$}}

\put(95,102){\makebox(35,8)[l]{${\bf v}_1$}}
\put(105,55){\makebox(35,8)[l]{${\bf v}_2$}}
\put(97,75){\makebox(35,8)[l]{${\bf c}$}}
\put(53.5,59){\makebox(35,8)[l]{${\bf u}$}}

\put(93.00, 85.00){\circle *{1.2}}
\put(92.78, 89.18){\circle *{1.2}}
\put(92.13, 93.32){\circle *{1.2}}
\put(91.04, 97.36){\circle *{1.2}}
\put(89.54, 101.27){\circle *{1.2}}
\put(87.64, 105.00){\circle *{1.2}}
\put(85.36, 108.51){\circle *{1.2}}
\put(82.73, 111.77){\circle *{1.2}}
\put(79.77, 114.73){\circle *{1.2}}
\put(76.51, 117.36){\circle *{1.2}}
\put(73.00, 119.64){\circle *{1.2}}
\put(69.27, 121.54){\circle *{1.2}}
\put(65.36, 123.04){\circle *{1.2}}
\put(61.32, 124.13){\circle *{1.2}}
\put(57.18, 124.78){\circle *{1.2}}
\put(53.00, 125.00){\circle *{1.2}}
\put(48.82, 124.78){\circle *{1.2}}
\put(44.68, 124.13){\circle *{1.2}}
\put(40.64, 123.04){\circle *{1.2}}
\put(36.73, 121.54){\circle *{1.2}}
\put(33.00, 119.64){\circle *{1.2}}
\put(29.49, 117.36){\circle *{1.2}}
\put(26.23, 114.73){\circle *{1.2}}
\put(23.27, 111.77){\circle *{1.2}}
\put(20.64, 108.51){\circle *{1.2}}
\put(18.36, 105.00){\circle *{1.2}}
\put(16.46, 101.27){\circle *{1.2}}
\put(14.96, 97.36){\circle *{1.2}}
\put(13.87, 93.32){\circle *{1.2}}
\put(13.22, 89.18){\circle *{1.2}}
\put(13.00, 85.00){\circle *{1.2}}
\put(13.22, 80.82){\circle *{1.2}}
\put(13.87, 76.68){\circle *{1.2}}
\put(14.96, 72.64){\circle *{1.2}}
\put(16.46, 68.73){\circle *{1.2}}
\put(18.36, 65.00){\circle *{1.2}}
\put(20.64, 61.49){\circle *{1.2}}
\put(23.27, 58.23){\circle *{1.2}}
\put(26.23, 55.27){\circle *{1.2}}
\put(29.49, 52.64){\circle *{1.2}}
\put(33.00, 50.36){\circle *{1.2}}
\put(36.73, 48.46){\circle *{1.2}}
\put(40.64, 46.96){\circle *{1.2}}
\put(44.68, 45.87){\circle *{1.2}}
\put(48.82, 45.22){\circle *{1.2}}
\put(53.00, 45.00){\circle *{1.2}}
\put(57.18, 45.22){\circle *{1.2}}
\put(61.32, 45.87){\circle *{1.2}}
\put(65.36, 46.96){\circle *{1.2}}
\put(69.27, 48.46){\circle *{1.2}}
\put(73.00, 50.36){\circle *{1.2}}
\put(76.51, 52.64){\circle *{1.2}}
\put(79.77, 55.27){\circle *{1.2}}
\put(82.73, 58.23){\circle *{1.2}}
\put(85.36, 61.49){\circle *{1.2}}
\put(87.64, 65.00){\circle *{1.2}}
\put(89.54, 68.73){\circle *{1.2}}
\put(91.04, 72.64){\circle *{1.2}}
\put(92.13, 76.68){\circle *{1.2}}
\put(92.78, 80.82){\circle *{1.2}}

\put(298.00, 85.00){\circle *{1.2}}
\put(297.78, 89.18){\circle *{1.2}}
\put(297.13, 93.32){\circle *{1.2}}
\put(296.04, 97.36){\circle *{1.2}}
\put(294.54, 101.27){\circle *{1.2}}
\put(292.64, 105.00){\circle *{1.2}}
\put(290.36, 108.51){\circle *{1.2}}
\put(287.73, 111.77){\circle *{1.2}}
\put(284.77, 114.73){\circle *{1.2}}
\put(281.51, 117.36){\circle *{1.2}}
\put(278.00, 119.64){\circle *{1.2}}
\put(274.27, 121.54){\circle *{1.2}}
\put(270.36, 123.04){\circle *{1.2}}
\put(266.32, 124.13){\circle *{1.2}}
\put(262.18, 124.78){\circle *{1.2}}
\put(258.00, 125.00){\circle *{1.2}}
\put(253.82, 124.78){\circle *{1.2}}
\put(249.68, 124.13){\circle *{1.2}}
\put(245.64, 123.04){\circle *{1.2}}
\put(241.73, 121.54){\circle *{1.2}}
\put(238.00, 119.64){\circle *{1.2}}
\put(234.49, 117.36){\circle *{1.2}}
\put(231.23, 114.73){\circle *{1.2}}
\put(228.27, 111.77){\circle *{1.2}}
\put(225.64, 108.51){\circle *{1.2}}
\put(223.36, 105.00){\circle *{1.2}}
\put(221.46, 101.27){\circle *{1.2}}
\put(219.96, 97.36){\circle *{1.2}}
\put(218.87, 93.32){\circle *{1.2}}
\put(218.22, 89.18){\circle *{1.2}}
\put(218.00, 85.00){\circle *{1.2}}
\put(218.22, 80.82){\circle *{1.2}}
\put(218.87, 76.68){\circle *{1.2}}
\put(219.96, 72.64){\circle *{1.2}}
\put(221.46, 68.73){\circle *{1.2}}
\put(223.36, 65.00){\circle *{1.2}}
\put(225.64, 61.49){\circle *{1.2}}
\put(228.27, 58.23){\circle *{1.2}}
\put(231.23, 55.27){\circle *{1.2}}
\put(234.49, 52.64){\circle *{1.2}}
\put(238.00, 50.36){\circle *{1.2}}
\put(241.73, 48.46){\circle *{1.2}}
\put(245.64, 46.96){\circle *{1.2}}
\put(249.68, 45.87){\circle *{1.2}}
\put(253.82, 45.22){\circle *{1.2}}
\put(258.00, 45.00){\circle *{1.2}}
\put(262.18, 45.22){\circle *{1.2}}
\put(266.32, 45.87){\circle *{1.2}}
\put(270.36, 46.96){\circle *{1.2}}
\put(274.27, 48.46){\circle *{1.2}}
\put(278.00, 50.36){\circle *{1.2}}
\put(281.51, 52.64){\circle *{1.2}}
\put(284.77, 55.27){\circle *{1.2}}
\put(287.73, 58.23){\circle *{1.2}}
\put(290.36, 61.49){\circle *{1.2}}
\put(292.64, 65.00){\circle *{1.2}}
\put(294.54, 68.73){\circle *{1.2}}
\put(296.04, 72.64){\circle *{1.2}}
\put(297.13, 76.68){\circle *{1.2}}
\put(297.78, 80.82){\circle *{1.2}}

\end{picture}
\vskip 10pt
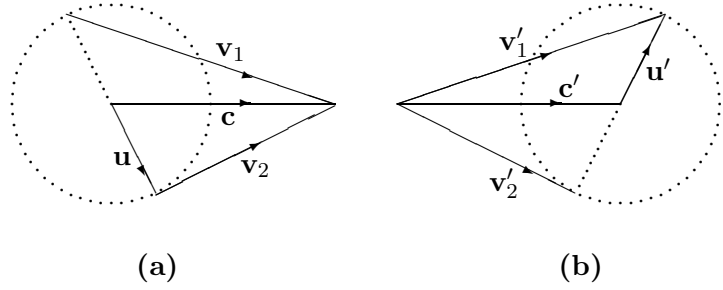
\noindent {\bf Figure 3}

\setlength{\unitlength}{0.013in}
\begin{picture}(200,130)

\put(63,0){\makebox(35,8)[l]{\bf (a)}}
\put(233,0){\makebox(35,8)[l]{\bf (b)}}

\put(196,36.5){\oval(5,10){}}
\put(196,35){\line(1,0){40}}
\put(196,38){\line(1,0){41}}
\put(236,35){\vector(1,2){21}}
\put(236,38){\vector(2,3){24}}
\put(180,42){\makebox(35,8)[l]{$dS$}}
\put(252,83){\makebox(35,8)[l]{$d\Omega_{{\bf u}^\prime}$}}
\multiput(241,36.5)(3.5,0){10}{\circle*{1.2}}
\put(276,31.5){\makebox(35,8)[l]{$\Omega_{\bf u}$}}

\put(32,35){\vector(1,0){48}}
\put(32,35){\vector(2,3){27}}
\multiput(77,35)(3.5,0){7}{\circle*{1.2}}
\multiput(56,71)(1.8,2.7){8}{\circle*{1.2}}
\put(52,24){\makebox(35,8)[l]{${\bf u}$}}
\put(102,30){\makebox(35,8)[l]{$\Omega_{\bf u}$}}
\put(35,56){\makebox(35,8)[l]{${\bf u}^\prime$}}
\put(72,95){\makebox(35,8)[l]{$\Omega_{{\bf u}^\prime}  $}}

\end{picture}

\newpage
\vspace{5pt}
\noindent{\bf Figure 4}

\setlength{\unitlength}{0.014in} 
\begin{picture}(200,145)

\put(59,42){\makebox(10,8)[l]{${\bf r}$}}

\put(138,36){\makebox(10,8)[l]{${\bf v}^\prime$}}
\put(125,83){\makebox(10,8)[l]{${\bf v}^\prime$}}

\put(86,58){\makebox(10,8)[l]{${\bf v}$}}
\put(122,62){\makebox(10,8)[l]{${\bf v}$}}

\put(60,52){\line(1,0){110}}
\put(60,52){\line(3,2){90}}
\put(60,52){\line(-1,0){20}}
\put(60,52){\line(-3,-2){16}}
\put(26,42){\makebox(10,8)[l]{$\Delta\Omega$}}
\put(45,69){\makebox(10,8)[l]{$-\Delta\Omega$}}

\put(72,57){\line(-1,1){10}}
\put(132,116){\circle{5}}
\put(172,78){\circle{5}}
\put(132,116){\vector(-1,-3){12}}
\put(120,80){\vector(-2,-1){28}}
\put(172,78){\vector(-4,-1){42}}
\multiput(156,73)(-1.3,-2.6){10}{\circle*{1.2}}
\put(143,47){\vector(-1,-2){1}}

\end{picture}

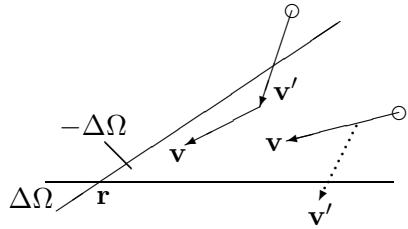
\noindent {\bf Figure 5}          

\setlength{\unitlength}{0.013in} 

\begin{picture}(200,150)

\put(63,15){\makebox(8,8)[l]{\bf (a)}}
\put(233,15){\makebox(8,8)[l]{\bf (b)}}

\put(76,55){\makebox(35,8)[l]{$\Delta{\bf r}$}}
\put(250,55){\makebox(35,8)[l]{$\Delta{\bf r}$}}
\put(64,55){\framebox(10,10){}}
\put(69,60){\line(1,4){16}}
\put(69,60){\line(-1,4){16}}

\put(76,75){\makebox(35,8)[l]{$-\Delta\Omega$}}
\multiput(69,60)(0,-2){10}{\circle*{1.2}}
\put(69,45){\vector(0,-1){5}}
\put(60,35){\makebox(35,8)[l]{\bf v}}
\put(236,55){\framebox(10,10){}}
\put(241,60){\line(1,4){16}}
\put(241,60){\line(-1,4){16}}
\multiput(246,60)(0.75,3){21}{\circle*{1.2}}
\multiput(236,60)(-0.75,3){21}{\circle*{1.2}}
\end{picture}

\vskip 10pt

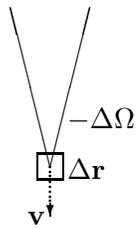
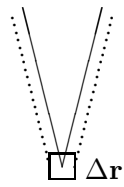
\noindent{\bf Figure 6}

\setlength{\unitlength}{0.013in}
\begin{picture}(300,140)

\put(63,5){\makebox(8,8)[l]{\bf (a)}}
\put(233,5){\makebox(8,8)[l]{\bf (b)}}

\put(67,31){\makebox(8,8)[l]{${\bf v}_1$}}
\put(48,60){\makebox(8,8)[l]{${\bf v}$}}
\put(92,78){\makebox(8,8)[c]{${\bf v}^\prime$}}
\put(98,50){\makebox(8,8)[c]{${\bf v}_1^\prime$}}
\put(8,35){\makebox(8,8)[c]{\bf r}}
\put(10,45){\line(1,0){90}}
\put(10,45){\line(1,1){55}}
\put(178,45){\line(1,0){90}}
\put(178,45){\line(1,1){55}}

\put(78,65){\circle*{3}}
 
 \put(71,64){\vector(-3,-1){35}}
\put(93,104){\vector(-1,-3){11}}
\put(89,105.5){\vector(-1,-3){11}}
\multiput(88.5,62.5)(2.5,0){15}{\circle*{1}}
\multiput(88.5,66.5)(2.5,0){15}{\circle*{1}}
\multiput(75.5,59)(-1.2,-2.4){13}{\circle*{1}}
\put(87,62.5){\vector(-1,0){1}}
\put(87,66.5){\vector(-1,0){1}}
\put(60.5,29){\vector(-1,-2){1}}

\put(176,35){\makebox(8,8)[c]{\bf r}}
\put(248,65){\circle*{3}}

\put(258,66){\vector(-1,0){1}}
\put(257.8,62){\vector(-4,1){1}}
\multiput(259,66)(2.5,0){12}{\circle*{1}}
\multiput(258.8,62)(2.5,-0.30){12}{\circle*{1}}
\put(238,61){\vector(-3,-1){28}}
\put(237,64){\vector(-4,-1){29}}
\put(258,110){\vector(-1,-4){8.5}}
\put(264,109){\vector(-1,-3){11.2}}

\multiput(245,54.5)(-0.8,-2){13}{\circle*{1}}
\multiput(242.5,56)(-1.2,-2){12}{\circle*{1}}
\put(236,32){\vector(-1,-4){1}}
\put(229,33.5){\vector(-1,-2){1}}

\put(272,92){\makebox(8,8)[c]{$f({\bf v}^\prime)$}}
\put(282,70){\makebox(8,8)[c]{$f({\bf v}_1^\prime)$}}
\put(214,65){\makebox(8,8)[c]{$f(\bf v)$}}
\put(238,27){\makebox(8,8)[l]{$f({\bf v}_1)$}}
\end{picture}

\newpage
\vspace{5pt}
\noindent {\bf Figure 7}

\setlength{\unitlength}{0.015in}
\begin{picture}(200,130)
\put(58,5){\makebox(0,8)[c]{$({\bf a})$}}
\put(148,5){\makebox(0,8)[c]{$({\bf b})$}}
\put(238,5){\makebox(0,8)[c]{$({\bf c})$}}

\put(58,25){\line(1,4){10}}
\put(58,25){\line(-1,4){10}}
\multiput(58,90)(0.25,-3){15}{\circle*{1.2}}
\multiput(58,90)(-0.25,-3){15}{\circle*{1.2}}
\put(51,22){\makebox(35,8)[l]{${\bf r}$}}
\put(65,40){\makebox(35,8)[l]{$-\Delta\Omega$}}
\put(62,75){\makebox(35,8)[l]{$\Delta\Omega_0$}}
\put(50,93){\makebox(35,8)[l]{${\bf r}_0$}}
\put(58,35){\line(4,1){14}}

\put(143,25){\framebox(10,10){}}
\multiput(148,90)(0.25,-3){19}{\circle*{1.2}}
\multiput(148,90)(-0.25,-3){19}{\circle*{1.2}}
\put(156,28){\makebox(35,8)[l]{$\Delta {\bf r}$}}
\put(152,75){\makebox(35,8)[l]{$\Delta\Omega_0$}}
\put(140,93){\makebox(35,8)[l]{${\bf r}_0$}}

\put(234,32){\framebox(8,8){}}
\multiput(238,90)(0.25,-3){17}{\circle*{1.2}}
\multiput(238,90)(-0.25,-3){17}{\circle*{1.2}}
\put(238,90){\line(0,-1){52}}
\put(238.5,35){\vector(0,-1){1}}
\put(242,75){\makebox(35,8)[l]{$\Delta\Omega_0$}}
\put(246,32){\makebox(35,8)[l]{$\Delta {\bf v}$}}
\put(238,23){\makebox(0,8)[c]{${\bf v}$}}
\end{picture}

\vspace{20pt}
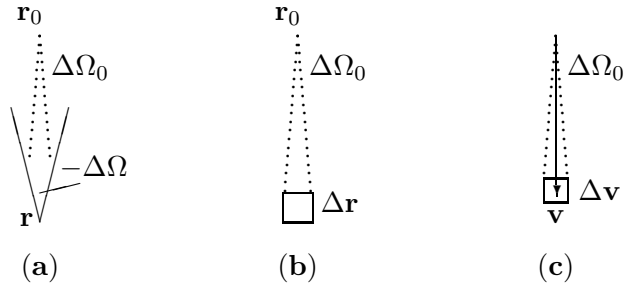
\noindent{\bf Figure 8}

\setlength{\unitlength}{0.013in}
\begin{picture}(200,135)

\put(135,93){\makebox(35,8)[l]{${\bf c}$}}
\put(190,50){\makebox(35,8)[l]{$\bf u$}}
\put(220,90){\vector(-1,-1){60}}
\put(70,90){\line(1,0){150}}
\put(217.5,90.5){\vector(1,0){1}}
\multiput(70,90)(3.0,-2.3){32}{\circle*{1.2}}
\multiput(70,90)(3.0,-1.7){35}{\circle*{1.2}}
\multiput(163,18.7)(1.6,2.4){6}{\circle*{1.2}}
\multiput(154,42.4)(-1.6,-2.4){5}{\circle*{1.2}}
\put(70,60){\makebox(35,8)[l]{$\Delta\Omega_0$}}
\put(140,14){\makebox(35,8)[l]{$\Delta v$}}

\put(143.22,53.87){\circle*{1}}
\put(144.80,50.69){\circle*{1}}
\put(146.52,47.57){\circle*{1}}
\put(148.36,44.53){\circle*{1}}
\put(150.32,41.57){\circle*{1}}
\put(152.41,38.70){\circle*{1}}
\put(154.62,35.91){\circle*{1}}
\put(156.94,33.22){\circle*{1}}
\put(159.38,30.63){\circle*{1}}
\put(161.91,28.15){\circle*{1}}
\put(164.56,25.77){\circle*{1}}
\put(167.29,23.50){\circle*{1}}
\put(170.13,21.35){\circle*{1}}
\put(173.04,19.32){\circle*{1}}
\put(176.04,17.42){\circle*{1}}
\put(179.12,15.64){\circle*{1}}
\put(182.27,14.00){\circle*{1}}
\put(185.49,12.48){\circle*{1}}

\end{picture}

\vspace{15pt}
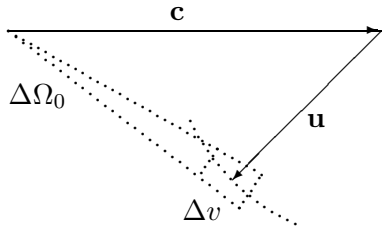
\noindent{\bf Figure 9}

\setlength{\unitlength}{0.013in}
\begin{picture}(200,130)
\put(100,15){\makebox(35,8)[l]{\bf (a)}}
\put(260,15){\makebox(35,8)[l]{\bf (b)}}
\put(240,90){\line(1,-1){40}}
\put(279,49.5){\vector(1,-1){1}}
\multiput(280.5,50)(3,-3){5}{\circle *{1.2}}
\put(295,30){\makebox(35,8)[l]{$\Omega_{{\bf u}^\prime}$}}
\put(258,74){\makebox(35,8)[l]{${\bf u}^\prime$}}

\put(110,93){\makebox(35,8)[l]{${\bf c}$}}
\put(143,63){\makebox(35,8)[l]{$\bf u$}}
\put(82,60){\makebox(35,8)[l]{$\bf v$}}
\put(143,30){\makebox(35,8)[l]{${\bf r}-{\bf r}_0$}}
\put(160,90){\vector(-1,-1){40}}
\put(60,90){\line(1,0){100}}
\put(158.5,90){\vector(1,0){1}}
\put(60,90){\line(3,-2){60}}
\put(118,50.5){\vector(3,-2){1}}
\multiput(120,50)(3,-2){6}{\circle*{1.2}}
\put(134,39){\vector(3,-2){5}}
\end{picture}

\newpage
\vspace{5pt}

\noindent {\bf Figure 10}

\setlength{\unitlength}{0.013in} 
\begin{picture}(200,145)
\put(67,40){\makebox(10,8)[l]{{\bf r}}}
\put(180,62){\oval(20,36){}}
\put(190,72){\line(2,1){20}}
\put(190,72){\vector(2,-1){20}}
\put(180,62){\line(1,0){10}}
\put(180,57){\line(1,0){10}}
\put(180,67){\line(1,0){10}}
\put(180,72){\line(1,0){8}}
\put(180,52){\line(1,0){8}}
\put(70,52){\line(1,0){100}}
\put(170,72){\vector(-4,-1){20}}
\put(130,72){\vector(-3,-1){40}}
\put(109,65){\vector(-2,1){20}}
\put(190,92){\vector(-3,-1){20}}
\put(190,92){\circle *{3}}
\put(70,52){\line(3,2){120}}
\put(180,107){\vector(-2,-1){30}}
\put(180,107){\line(0,2){25}}
\put(70,52){\circle *{3}}
\end{picture}
\vspace{5pt}

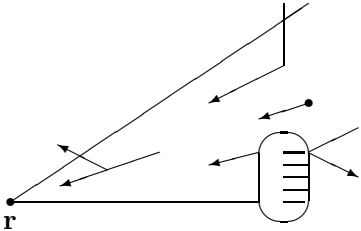
\noindent {\bf Figure 11}

\setlength{\unitlength}{0.013in} 
\begin{picture}(200,165)
\put(63,15){\makebox(35,8)[l]{\bf (a)}}
\put(233,15){\makebox(35,8)[l]{\bf (b)}}

\put(112,42){\makebox(35,8)[l]{${\bf r}$}}

\multiput(45,85)(3,0){20}{\line(0,1){5}}

\multiput(20,40)(10,0){2}{\line(0,1){45}}
\multiput(20,90)(10,0){2}{\line(0,1){45}}
\multiput(20,60)(0,5){12}{\line(1,0){10}}
\put(35,90){\vector(1,0){80}}
\put(35,85){\vector(1,0){80}}
\put(98,55){\vector(1,-1){1}}
\put(105,55){\vector(1,-2){1}}
\put(70,85){\line(1,-1){40}}
\put(90,85){\line(1,-2){20}}
\multiput(45,85)(3,0){20}{\line(0,1){5}}

\multiput(190,40)(10,0){2}{\line(0,1){45}}
\multiput(190,90)(10,0){2}{\line(0,1){45}}
\multiput(190,60)(0,5){12}{\line(1,0){10}}
\put(205,87.5){\vector(1,0){80}}

\put(259,65){\vector(1,-1){1}}
\put(237.5,87.5){\line(1,-1){56}}
\multiput(280,87.5)(0,-3){14}{\circle*{1}}
\put(283,67){\makebox(35,8)[l]{$r_\perp$}}
\put(249,60){\makebox(35,8)[l]{${\bf v}$}}
\put(291,36){\makebox(35,8)[l]{$\theta$}}
\put(280,45){\line(1,0){18}}
\end{picture}

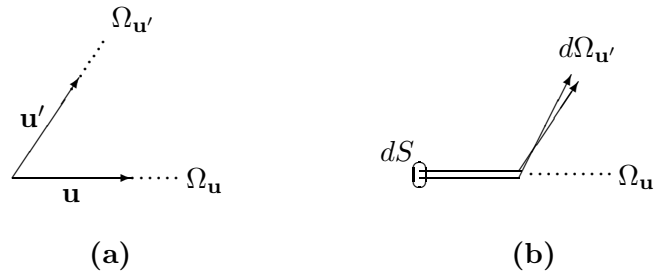
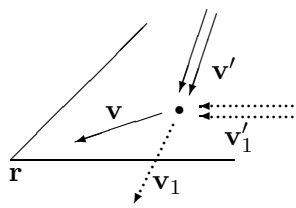
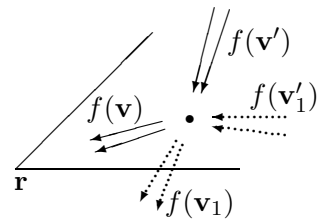
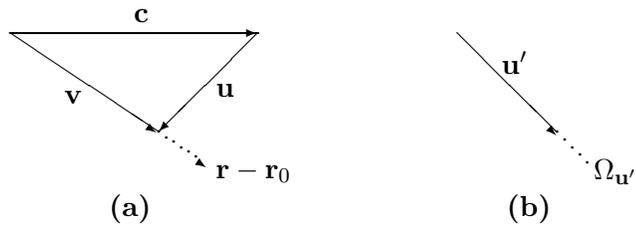
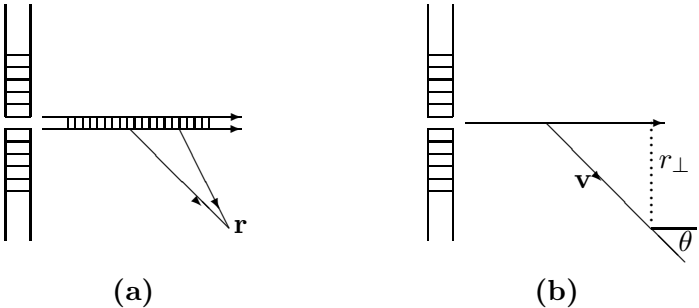
\end{document}